\documentclass[aip, amsmath, amssymb, reprint, prx, nofootinbib]{revtex4-2}

\usepackage{amsfonts}
\usepackage{amsmath}
\usepackage{amssymb}
\usepackage{filecontents}
\usepackage{array}   % table making
\usepackage{bm}      % bold math
\usepackage{braket}
\usepackage{xcolor}
\usepackage{dcolumn} % Align table columns on decimal point
\usepackage[T1]{fontenc}
\usepackage{graphicx}
\usepackage[utf8]{inputenc}
\usepackage{mathptmx}
\usepackage{tabularx}
\usepackage{url}
\usepackage{xr-hyper}
\usepackage[bookmarks=false,colorlinks,citecolor=blue]{hyperref}

\usepackage{amsthm,enumitem}
\usepackage{caption}
\usepackage{subcaption}

\makeatletter
\newcommand*{\addFileDependency}[1]{
  \typeout{(#1)}
  \@addtofilelist{#1}
  \IfFileExists{#1}{}{\typeout{No file #1.}}
}
\makeatother

\newcommand*{\myexternaldocument}[1]{
    \externaldocument{#1}
    \addFileDependency{#1.tex}
    \addFileDependency{#1.aux}
}
\myexternaldocument{supplement}
\begin{document}
\include{MyCommand}
\setlength{\tabcolsep}{5pt} % for the horizontal padding
\renewcommand{\arraystretch}{1.1}

\title{A many-body characterization of the fundamental gap in monolayer CrI$_3$}

\author{Daniel Staros\textsuperscript{\textdagger}}
\affiliation{Department of Chemistry, Brown University, Providence, Rhode Island 02912, USA}
\author{Abdulgani Annaberdiyev\textsuperscript{\textdagger}}
\affiliation{\mbox{Center for Nanophase Materials Sciences Division, Oak Ridge National Laboratory, Oak Ridge, Tennessee 37831, USA}}
\author{Kevin Gasperich}
\affiliation{\mbox{Computational Science Division, Argonne National Laboratory, Lemont, Illinois 60439, USA}\\
\mbox{Current: Qubit Pharmaceuticals, Incubateur Paris Biotech Santé, 24 rue du Faubourg Saint Jacques, 75014 Paris, France}}
\author{Anouar Benali}
\affiliation{\mbox{Computational Science Division, Argonne National Laboratory, Lemont, Illinois 60439, USA}\\
\mbox{Current: Qubit Pharmaceuticals, Incubateur Paris Biotech Santé, 24 rue du Faubourg Saint Jacques, 75014 Paris, France}}

\author{Panchapakesan Ganesh}
\email{ganeshp@ornl.gov}
\affiliation{\mbox{Center for Nanophase Materials Sciences Division, Oak Ridge National Laboratory, Oak Ridge, Tennessee 37831, USA}}

\author{Brenda Rubenstein}
\email{brenda\_rubenstein@brown.edu}
\affiliation{Department of Chemistry, Brown University, Providence, Rhode Island 02912, USA}

%\date{\today}

\begin{abstract}
The many-body fixed-node and fixed-phase spin-orbit Diffusion Monte Carlo (DMC) methods are applied to accurately predict the fundamental gap of monolayer CrI$_3$ -- the first experimentally-realized 2D material with intrinsic magnetism.
The fundamental gap obtained, 2.9(1)~eV, agrees well with the highest peak in optical spectroscopy measurements and a previous $GW$ result.
We numerically show that as expected in DMC the same value of the fundamental gap is obtained in the thermodynamic limit using both neutral promotions and the standard quasiparticle definition of the gap based on the ionization potential and electron affinity.  Additional analysis of the differences between density matrices formed in different bases using configuration interaction calculations explains why a single-reference trial wave function can produce an accurate excitation.
We find that accounting for electron correlation is more crucial than accounting for spin-orbit effects in determining the fundamental gap.
These results highlight how DMC can be used to benchmark 2D material physics and emphasize the importance of using beyond-DFT methods for studying 2D materials.
\end{abstract}

%\pacs{}
\maketitle
%\pagenumbering{arabic}

\section{Introduction}
\label{sec:intro}
\footnote{Notice:  This manuscript has been authored by UT-Battelle, LLC, under contract DE-AC05-00OR22725 with the US Department of Energy (DOE). The US government retains and the publisher, by accepting the article for publication, acknowledges that the US government retains a nonexclusive, paid-up, irrevocable, worldwide license to publish or reproduce the published form of this manuscript, or allow others to do so, for US government purposes. DOE will provide public access to these results of federally sponsored research in accordance with the DOE Public Access Plan (http://energy.gov/downloads/doe-public-access-plan).}
Magnetic two-dimensional (2D) materials are of great interest in condensed matter physics due to their potential applications in spintronics, quantum computing, and neuromorphic computing.~\cite{park_opportunities_2016, wang_tunneling_2018, ningrum_recent_2020, liu_recent_2023, fu_tunneling_2024, elahi_recent_2024, zhang_2d_2024, Staros_npj2024} However, 2D materials can be challenging to model with the most commonly employed Kohn-Sham Density Functional Theory (KS-DFT) methods, partly due to the difficulties associated with accurately capturing electron correlation effects using DFT.
Since these systems often display strong excitonic character,~\cite{Wu_natcomm2019} a particular point of interest for condensed matter theory is accurately determining the fundamental gaps of these materials, which in turn inform their exciton binding energies.

This work focuses on chromium triiodide (CrI$_3$), which in its bulk form, is a layered van der Waals material consisting of a honeycomb lattice of Cr atoms, each surrounded by an octahedra of iodine atoms. CrI$_3$ is an out-of-plane ferromagnetic system, due to the nominal $3+$ charge on the chromium, which corresponds to an intrinsic magnetic moment of $3~\mu_\mathrm{B}$.~\cite{McGuire_ChemMater}
Remarkably, it was demonstrated that its intrinsic magnetism is preserved upon thinning the bulk, even down to the monolayer (ML) limit.~\cite{Huang_natlett2017}
Specifically, ML CrI$_3$ was experimentally synthesized using micromechanical exfoliation techniques and found to have a Curie temperature of 45~K,~\cite{Huang_natlett2017} making it the first fabricated 2D material that displays intrinsic magnetism. 

Following the experimental synthesis and characterization of ML CrI$_3$,~\cite{Huang_natlett2017, Seyler_np2018} many computational studies have examined its electronic structure. The most common method employed has been KS-DFT, although beyond-DFT methods such as the Green's function-based $GW$ method and its various flavors,~\cite{Wu_natcomm2019, Sanchez_jmcc2020, Schilfgaarde_prb2021} dynamical mean-field theory (DMFT),~\cite{kvashnin_dynamical_2022} and quantum Monte Carlo (QMC)~\cite{Staros_jcp2022, wines_systematic_2023} methods have been applied. These studies have examined CrI$_{3}$'s various electronic structure properties such as its geometry, charge density, magnetic moments, excitonic gap, and quasiparticle or fundamental gaps. After some iteration in the community, ground-state quantities, such as the geometry and magnetic moments, now show reasonable agreement amongst these beyond-DFT methods, as well as with experimental observations.

However, the fundamental gap of ML CrI$_3$ has been a point of controversy.
The KS-DFT method with various exchange-correlation (XC) functionals has predicted the gap to be 0.82\cite{Wu_natcomm2019} to 1.06 eV,~\cite{Schilfgaarde_prb2021} while $GW$ methods and their variants have produced results in the range of 2.59\cite{Wu_natcomm2019} to 3.25\cite{Schilfgaarde_prb2021} eV.
Experimentally, the optical reflection spectrum shows multiple excitonic peaks and a dominant peak at about 2.7~eV.~\cite{Seyler_np2018}
As of writing this article, a direct experimental measurement of the fundamental gap using angle-resolved photoemission spectroscopy (ARPES) is unavailable.

This picture is further complicated if a comparison is made~\cite{Kutepov_prm2021} with the fundamental gap of bulk CrI$_3$. The ARPES measurements for the bulk have yielded a fundamental gap of 1.3~eV,~\cite{kundu_valence_2020} which is in better agreement with the KS-DFT values of the electronic band gap in ML CrI$_3$. However, the decreased electronic screening of ML CrI$_3$ relative to the bulk should be reflected in a larger fundamental gap.~\cite{Latini_prb2015} Evidently, GW methods qualitatively recover this expectation,~\cite{Wu_natcomm2019,Schilfgaarde_prb2021} but the significant disagreement regarding the fundamental gap among the methods necessitates a quantitative, beyond-perturbative determination of the fundamental gap in ML CrI$_3$.

In this work, we use the highly accurate Diffusion Monte Carlo (DMC) method to estimate the fundamental gap of ML CrI$_3$ and resolve this conundrum.
DMC is a many-body stochastic electronic-structure method that has been successfully applied to predicting gaps in molecules,~\cite{blunt_excited-state_2019} 1D materials,~\cite{mostaani_quasiparticle_2016} 2D materials,~\cite{frank_many-body_2019, dubeckyFundamentalGapFluorographene2020, Shin_PhysRevMats, wines_towards_2024} and solids.~\cite{meltonManybodyElectronicStructure2020}
In particular, we use two flavors of DMC: fixed-node DMC (FNMDC),~\cite{Foulkes_rmp2001} which is a scalar-relativistic framework, and the fixed-phase spin-orbit DMC (FPSODMC),~\cite{meltonQuantumMonteCarlo2016, meltonFixedNodeFixedPhaseApproximations2016, meltonSpinorbitInteractionsElectronic2016,meltonQuantumMonteCarlo2017,meltonProjectorQuantumMonte2019} which is a fully-relativistic treatment.
We find that both methods predict the fundamental gap to be $2.9 \pm 0.1$ eV,
despite significant differences in the KS-DFT gaps of their underlying trial wave functions, which is close to the experimental value~\cite{Seyler_np2018}.
This implies that the nature of the fundamental gap is mainly determined by the electron correlation effects.
Additional configuration interaction calculations are also presented to further characterize the many-body nature of the gap, which also provides a justification for the use of single-reference trial wave functions in DMC. Our work paves the way for using DMC to perform benchmark calculations of fundamental gaps in open-shell extended correlated 2D materials.

\section{Methods}
\label{sec:methods}

%\subsection{Structure and Pseudopotentials}
All calculations employ an 8-atom primitive cell of monolayer CrI$_3$ with an out-of-plane vacuum larger than 25~\r{A} to avoid spurious interactions among monolayer images.
This structure was previously optimized using FNDMC via a high-accuracy surrogate Hessian line-search method that was shown to agree with  experiment with an error of less than 0.5\%.~\cite{Staros_jcp2022} See the SI for the employed geometry.

We utilize pseudopotentials for Cr and I, with 14 and 7 valence electrons, respectively. The Cr pseudopotential which explicitly includes semicore states with 14 explicitly-treated electrons is more accurate than a simple valence Cr pseudopotential since fewer electrons are pseudized in the 14-valence case; transferability is not affected by this choice. Cr is always modeled using a scalar-relativistic (SR) potential, while I is modeled using SR and fully-relativistic (FR) pseudopotentials.
Specifically, we use an RRKJ-type potential for Cr,~\cite{krogel_pseudopotentials_2016} and an energy-consistent BFD pseudopotential for I.~\cite{burkatzki_energy-consistent_2007}
To validate these potentials, we tested them against correlation-consistent effective core potentials (ccECP), which are harder pseudopotentials optimized using correlated calculations.~\cite{kincaidCorrelationConsistentEffective2022, wang_new_2022}
We found that the differences between the two sets are negligible, and we can assume that the employed pseudopotentials are accurate enough for the purposes of this work (see the SI).

%\subsection{DFT Methods}
We use the open-source \texttt{Quantum ESPRESSO} package~\cite{Giannozzi_jpcm2009, Giannozzi_jpcm2017} to obtain initial DFT trial wave functions for our Diffusion Monte Carlo calculations.
These calculations are converged on a Monkhorst Pack \textbf{k}-grid of \mbox{14×14×1} and with a plane wave energy cutoff of 300~Ry.
All calculations were performed for the out-of-plane ferromagnetic configuration of ML CrI$_3$.
We use the PBE+$U$ exchange-correlation functional~\cite{Perdew_prl1996} with a Hubbard $U$ correction of 2 eV per Cr (denoted as PBE$+U_{2.0}$) in accordance with previous results showing that this $U$ value leads to a DFT description that best matches the QMC results.~\cite{Ichiba_prm2021,Staros_jcp2022}
In addition, the variational minimum of the FNDMC ground state energy was obtained for a $U \approx 2.0$~eV.\cite{Ichiba_prm2021,Staros_jcp2022,wines_systematic_2023}

%\subsection{QMC Methods}
All DMC calculations were carried out with the \texttt{QMCPACK} code~\cite{Kim_jpcm2018, Kent_jcp2020} and driven by \texttt{Nexus}~\cite{Krogel_CompPhysComm} workflows.
We use single-reference Slater-Jastrow trial wavefunctions built from PBE+$U_2$ eV plane wave Bloch orbitals.
In FNDMC, the trial wave function can be written as
\begin{equation}
    \Psi_{\rm T}(\textbf{R}) = D^{\uparrow}(\textbf{R}) D^{\downarrow}(\textbf{R}) e^{J(\textbf{R})},
\end{equation}
where $D^{\uparrow}$ and $D^{\downarrow}$ are up and down determinants, respectively.
The $e^J$ represents the symmetric Jastrow factor, which includes one-, two-, and three-body terms.
The spin multiplicity of $2S+1 = 7$ per primitive cell (or equivalently, 3~$\mu_B$ per Cr) is enforced in all supercells.

In FPSODMC, there is no spin up and down electron distinction, and the trial wave function can no longer be partitioned into up and down determinants.
Instead, it consists of a single determinant with spinor orbitals: 
\begin{equation}
    \Psi_{\rm T}(\textbf{R, s}) = D(\textbf{R, s}) e^{J(\textbf{R})}
\end{equation}
where $\mathbf{s}$ is a spin coordinate to be sampled in QMC.

Pseudopotential evaluation was performed using T-moves to minimize localization errors~\cite{Casula_JChemPhys,Dzubak_JChemPhys}.
All DMC calculations utilized a timestep of $d\tau=0.01$ Ha$^{-1}$ since this was previously shown to be sufficient for this system~\cite{Staros_jcp2022}.
We do not apply corrections to the kinetic energy,\cite{chiesa_finite-size_2006} or potential energy,\cite{fraser_finite-size_1996, williamson_elimination_1997, kent_finite-size_1999} or the resulting gaps\cite{Hunt_prb2018}, and report the raw quantities throughout the paper.

\section{Results}
\label{sec:results}

\begin{figure*}[!htbp]
\centering
\includegraphics[width=0.9\textwidth]{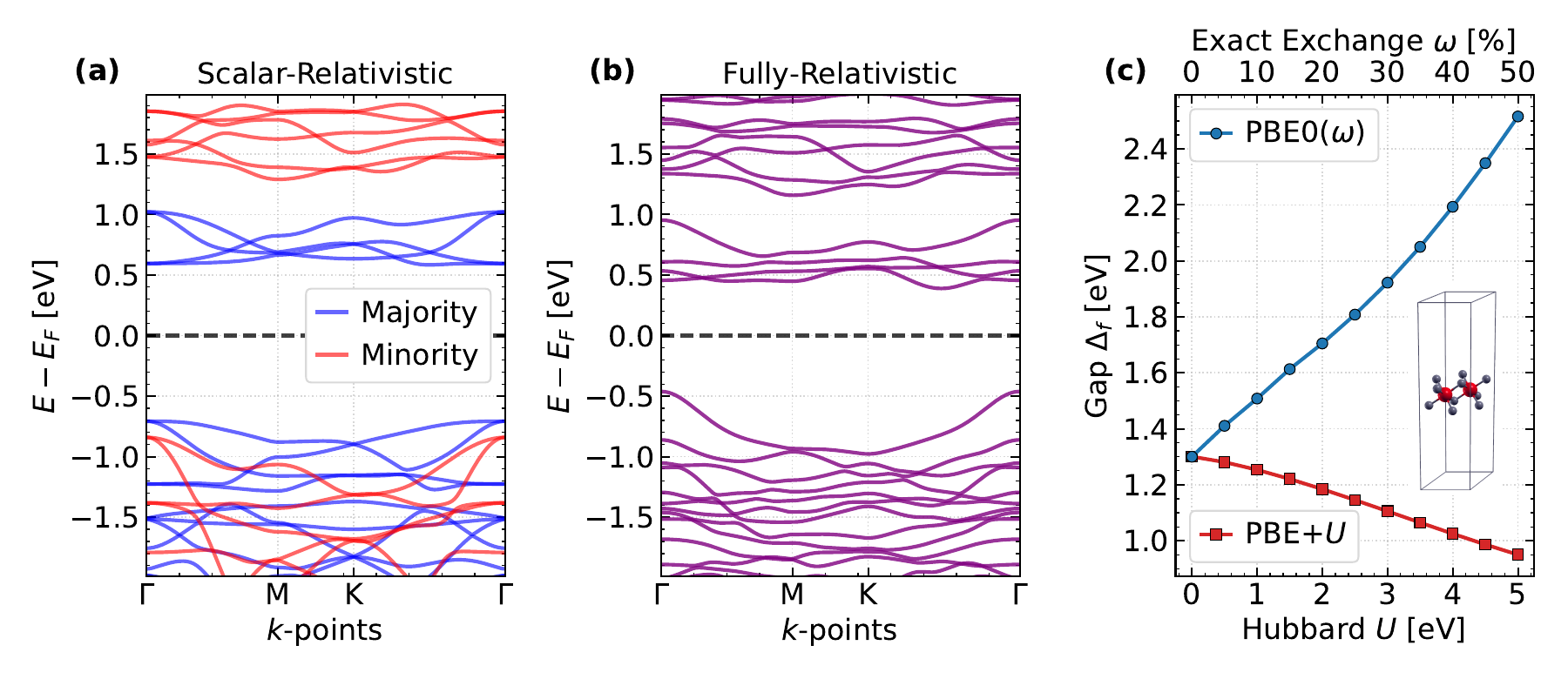}
\caption{
Electronic structure of ML CrI$_3$ using KS-DFT.
(a) Scalar-relativistic, collinear band structure using the PBE functional.
The majority (spin-up, blue) and minority (spin-down, red) channels are shown.
(b) Fully-relativistic, non-collinear band structure using the PBE functional.
(c) Scalar-relativistic gaps using PBE$+U$ and PBE0($\omega$).
The inset shows the primitive cell structure employed.
}
\label{fig:dft_res}
\end{figure*}

As mentioned previously, ML CrI$_3$ has a ferromagnetic ground magnetic state, which is confirmed by both experiments and theory.\cite{McGuire_ChemMater,Huang_natlett2017}
An approximate SR band structure obtained with the PBE XC functional is shown in Fig.~\ref{fig:dft_res}(a).
We can see that the fundamental gap consists of a spin-conserved promotion in the majority spin channel.
The valence band maximum (VBM) and the conduction band minimum (CBM) are very close to the $\Gamma$-point and the bands are flat near that region.
Therefore, we treat the fundamental gap as a $\Gamma_\mathrm{VBM} \rightarrow \Gamma_\mathrm{CBM}$ promotion.
Even with the inclusion of spin-orbit coupling (SOC), $\Gamma_\mathrm{VBM} \rightarrow \Gamma_\mathrm{CBM}$ approximates the true band gap well, despite significant changes in the electronic structure, Fig.~\ref{fig:dft_res}(b).
This minor approximation allows us to extrapolate using much smaller supercells in our QMC calculations.

In Fig.~\ref{fig:dft_res}(c), the SR fundamental gap $\Delta_f$ produced using PBE$+U$ and PBE0($\omega$)~\cite{Adamo_jms1999} with varying values of the Hubbard $U$ and exact exchange $\omega$ is shown. The PBE$+U$ gaps \textit{decrease} in value as also seen elsewhere\cite{Sanchez_jmcc2020} due to the increased splitting of the Cr 3$d$-manifold with larger $U$. Specifically, applying a larger $U$ pushes the occupied 3$d$-states deeper into the Fermi sea, resulting in the decreased screening and higher energy of the I 5$p$-states. In other words, within the KS-DFT formalism, the application of the Hubbard $U$ correction pulls the $\Delta_f$ down from the gap indicated by the highest peak in the optical reflectivity measurements (2.7 eV)~\cite{Seyler_np2018}, which is assumed to be the fundamental gap here.
Hybrid DFT using PBE0($\omega$) does increase the gaps as expected.
However, for an accurate $\Delta_f$, the exact exchange mixing must be set at an unusually high value of $\omega > 50\%$, when compared to the most commonly used XC functionals such as B3LYP~\cite{Becke_jcp1993} with $\omega = 20\%$, and PBE0 with $\omega = 25\%$.
The fact that the PBE$+U$ and PBE0($\omega$) gaps trend in opposite directions combined with the inaccurate gap values at commonly used $+U$ and $\omega$ values paints a clear picture of where KS-DFT struggles. These struggles call for using beyond-DFT methods, such as $GW$ or QMC, on this material.

The excited states of ML CrI$_3$ have been extensively studied using various $GW$ methods\cite{Wu_natcomm2019, Sanchez_jmcc2020, Schilfgaarde_prb2021}.
Here, we provide the complementary data from FNDMC and FPSODMC calculations of the fundamental gap.
To obtain $\Delta_f$, we employ two formally different expressions.
In one approach, the gap is calculated using the standard definition of the fundamental gap (or quasiparticle gap), which involves the calculation of neutral and charged systems with $N$, $N+1$, and $N-1$ electrons at the $\Gamma$-point:
\begin{equation}
    \begin{aligned}
    \label{eqn:def_qp}
    \Delta_{f}^{\rm qp}(\Gamma) & = \mathrm{IP} - \mathrm{EA} = \\
    & = \left[E_{N+1}(\Gamma)-E_N(\Gamma)\right] - \left[E_{N}(\Gamma)-E_{N-1}(\Gamma)\right],
    \end{aligned}
\end{equation}
where IP and EA stand for the ionization potential and electron affinity, respectively.
In the second approach, we use the neutral excitation of an electron from the VBM to CBM at the $\Gamma$-point:
\begin{equation}
    \label{eqn:def_ex}
    \Delta_{f}^{\rm opt}(\Gamma) = E^\dag_{N}(\Gamma)-E_N(\Gamma).
\end{equation}
In general, this expression is used to obtain the optical spectrum of the system, which might involve excitonic states.
Therefore, one might expect that the use of Eqn.~\ref{eqn:def_ex} is not justified for the calculation of fundamental gaps.
However, within the use of KS-DFT trial wavefunctions, which have translational symmetry, both definitions are expected to recover the true fundamental gap of the system in the thermodynamic limit (TDL), $\Delta_{f}^{\rm qp, TDL} = \Delta_{f}^{\rm opt, TDL}$ \cite{dubeckyFundamentalGapFluorographene2020}.
This can be intuitively understood since the excitonic states should have a localized character, whereas the promoted electron in Eqn.~\ref{eqn:def_ex} will occupy the delocalized CBM, and this character will persist to arbitrarily large supercells and the TDL.
As pointed out previously \cite{dubeckyFundamentalGapFluorographene2020}, a similar convergence can be seen in KS-DFT for $G = \rm IP - EA$ and $g = \varepsilon^{LU} - \varepsilon^{HO}$ (one-particle eigenvalues), resulting in $G = g$ at the TDL\cite{perdewUnderstandingBandGaps2017}.

\begin{figure}[!htbp]
\centering
\includegraphics[width=0.5\textwidth]{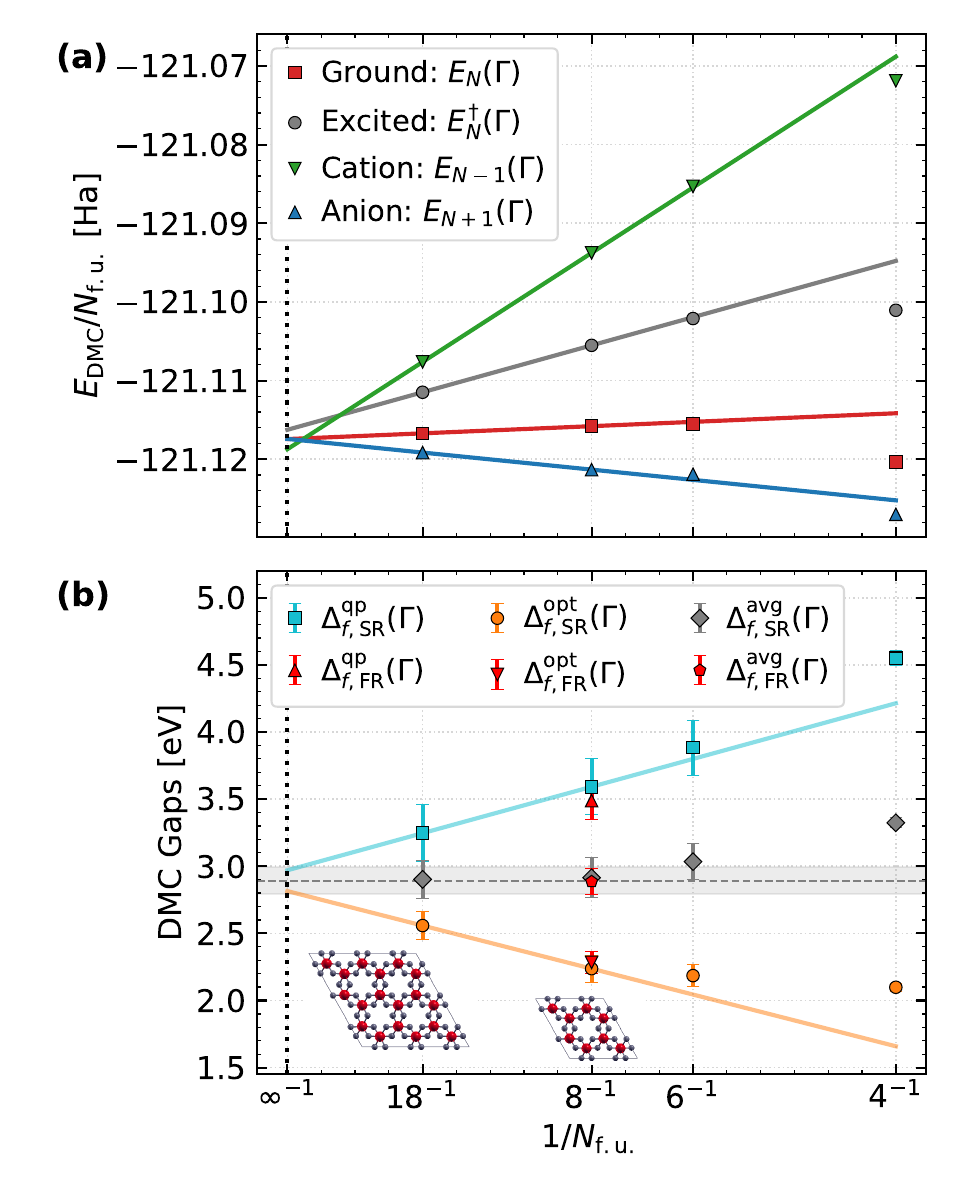}
\caption{
Finite-size scaling of DMC total energies and fundamental gaps for ML CrI$_3$.
(a) Scalar-relativistic FNDMC total energies per formula unit.
(b) DMC fundamental gaps using the $\Delta^{\rm qp}_{f}(\Gamma)$ (Eqn.~\ref{eqn:def_qp}) and $\Delta^{\rm opt}_{f}(\Gamma)$ (Eqn.~\ref{eqn:def_ex}) definitions.
$\Delta^{\rm avg}_{f}(\Gamma) = [\Delta^{\rm qp}_{f}(\Gamma) + \Delta^{\rm opt}_{f}(\Gamma)]/2$ is the average of the two definitions, and the dashed horizontal line and the gray envelope represent its extrapolated value and error in the TDL, respectively. 
Fundamental gaps from fully-relativistic FPSODMC are shown with red points and indicated by the ``FR'' label.
}
\label{fig:gaps}
\end{figure}

We use successively larger supercells to extrapolate to the TDL, which is shown in Fig.~\ref{fig:gaps}.
The data for $N_\mathrm{f.u.} = 4, 6, 8, 18$ formula units is shown ($N_\mathrm{f.u.} = 2$ is the primitive cell).
To obtain the TDL values for the total energies and gaps, we use a two-point linear extrapolation using the largest two supercells of $N_\mathrm{f.u.} = 8, 18$, which are the uniformly tiled $[2\times2\times1]$ and $[3\times3\times1]$ supercells of the primitive cell, respectively.

The convergence of the total energies with respect to the inverse cell size is shown in Fig.~\ref{fig:gaps}(a).
At the smallest cell size ($N_\mathrm{f.u.} = 4$), we see non-linear behavior in the energies and gaps, most likely due to shell-filling effects (i.e., unconverged kinetic energies)~\cite{Holzmann_PRB}.
At the larger supercell sizes ($N_\mathrm{f.u.} = 6, 8, 18$), the trends are consistent with linear behavior, as expected.
We obtain a reasonable agreement between the energies and gaps in the TDL using the ground, excited, cation, and anion states (with about 1~mHa deviations around the mean value), which is also expected since the energy per formula unit should not depend on $O(1)$ shifts at infinity.
This signifies that the supercells employed with $N_\mathrm{f.u.} = 8, 18$ are sufficiently large for the purposes of estimating $\Delta_f$ in the TDL.

The finite-size convergence of $\Delta_{f}^{\rm qp}(\Gamma)$ and $\Delta_{f}^{\rm opt}(\Gamma)$ is shown in Fig.~\ref{fig:gaps}(b).
We see that $\Delta_{f}^{\rm qp}(\Gamma)$ converges from above while $\Delta_{f}^{\rm opt}(\Gamma)$ converges from below.
Although having different signs for slopes is not generally guaranteed, a previous 2D calculation on GeSe~\cite{Shin_PhysRevMats} is in agreement with $\Delta_{f}^{\rm opt}$ converging from below $\Delta_{f}^{\rm qp}$.
In Fig.~\ref{fig:gaps}(b), the difference ($\Delta_{f}^{\rm qp}(\Gamma) - \Delta_{f}^{\rm opt}(\Gamma)$) is quite large at the smallest supercell size, on the order of a few eV, and consistently decreases as $N_\mathrm{f.u.}$ increases.
Upon extrapolation with the $N_\mathrm{f.u.} = 8, 18$ data points, the TDL values differ by about $0.16$~eV (and agree within the error bars).
We expect that this difference should go to zero with larger supercell extrapolations, although we cannot afford them at this point.
Therefore, we take the best FNDMC estimate in the TDL to be the average of the two approaches, $\Delta_{f}^{\rm qp} = 2.97$~eV and $\Delta_{f}^{\rm opt} = 2.81$~eV, with an error bar of $\pm 0.1$~eV, giving $\Delta_{f,\rm SR}^{\rm DMC} = 2.9(1)$~eV.

Since our FNDMC calculations used SR pseudopotentials and collinear trial wave functions, we now test for the possible contribution to the gap from spin-orbit coupling using the FPSODMC method.
The non-collinear FPSODMC calculations are more expensive than the collinear FNDMC calculations because they require complex trial wave functions and the sampling of one more additional spin degree of freedom~\cite{meltonQuantumMonteCarlo2016, meltonFixedNodeFixedPhaseApproximations2016, meltonSpinorbitInteractionsElectronic2016,meltonQuantumMonteCarlo2017,meltonProjectorQuantumMonte2019}.
Therefore, we were able to afford FPSODMC only for a [$2\times2\times1$] supercell, as shown by the red symbols in Fig.~\ref{fig:gaps}(b).
However, note that the average of the scalar-relativistic $\Delta_{f, \rm SR}^{\rm qp}(\Gamma)$ and $\Delta_{f, \rm SR}^{\rm opt}(\Gamma)$ values already agree with the TDL value at this size, Fig.~\ref{fig:gaps}(b).
Assuming that the same finite-size scaling applies to FR calculations, we use the average values of $\Delta_{f, \rm FR}^{\rm qp}(\Gamma)$ and $\Delta_{f, \rm FR}^{\rm opt}(\Gamma)$ for the [$2\times2\times1$] supercell to provide the $\Delta_{f,\rm FR}^{\rm DMC}$ estimate in the TDL.
As seen in Fig.~\ref{fig:gaps}(b), the  FPSODMC gaps obtained agree with the FNDMC gaps within the error bars.
This is despite the fact that the underlying PBE$+U_{2.0}$ functional predicts a decrease in $\Delta_f$ by about $0.5$~eV due to non-collinearity and SOC (shown in Table~\ref{tab:gaps}). This signifies that $\Delta_f$ is mainly determined by electron correlation effects, and the SOC plays a minor role that could have an effect up to a few tenths of an eV, which we cannot resolve within statistical error bars at this point. 

\begin{figure}[!htbp]
\centering
\includegraphics[width=0.5\textwidth]{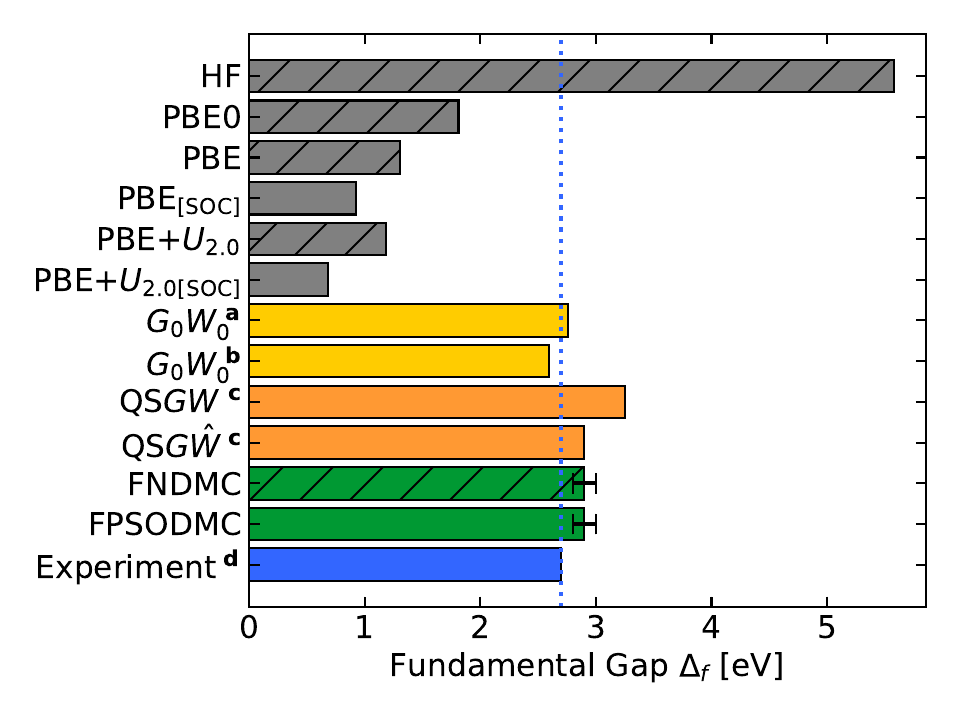}
\caption{
Summary of fundamental gaps of ML CrI$_3$.
Unrestricted Kohn Sham (UKS) DFT and Diffusion Monte Carlo (DMC) data are from the current study.
Superscripts represent other studes: $a$~\cite{Sanchez_jmcc2020}, $b$~\cite{Wu_natcomm2019}, $c$~\cite{Schilfgaarde_prb2021}, $d$~\cite{Seyler_np2018}.
The hatched data is scalar-relativistic, while the unhatched data is fully-relativistic.
The dotted vertical line is the experimental value as a guide to the eye.
}
\label{fig:all_gaps}
\end{figure}

We compare our DMC fundamental gaps with those obtained from various flavors of $GW$ in Table~\ref{tab:gaps}.
In addition, the table provides the $\Delta_f$ using common KS-DFT XC functionals and unrestricted Hartree-Fock (HF), which severely overestimates the gap.
These gaps are also visually summarized in Fig.~\ref{fig:all_gaps}.

\begin{table}[htbp!]
\centering
\caption{Fundamental gap values predicted in this work compared to those from other studies.
XC/Orbital represents the underlying exchange-correlation functional or orbital type used in the method.
Scalar-relativistic (SR) and fully-relativistic (FR) values are shown.
}
\begin{tabular}{llllr}
\hline
Method                      & XC/Orbital    & Rel. & $\Delta_f$ [eV] & Ref. \\
\hline
UHF                         &               & SR & 5.58   & \textit{this work} \\
UKS                         & PBE0          & SR & 1.81   & \textit{this work} \\
UKS                         & PBE           & SR & 1.30   & \textit{this work} \\
UKS                         & PBE           & FR & 0.92   & \textit{this work} \\
UKS                         & PBE$+U_{2.0}$ & SR & 1.18   & \textit{this work} \\
UKS                         & PBE$+U_{2.0}$ & FR & 0.68   & \textit{this work} \\
\\
FNDMC                       & PBE$+U_{2.0}$ & SR & 2.9(1) & \textit{this work} \\
FPSODMC                     & PBE$+U_{2.0}$ & FR & 2.9(1) & \textit{this work} \\
\\                            
$G_0W_0$                    & LDA$+U_{1.5}$ & FR & 2.76   & Ref.~\cite{Sanchez_jmcc2020} \\
$G_0W_0$                    & LDA$+U_{1.5}$ & FR & 2.59   & Ref.~\cite{Wu_natcomm2019} \\
\\                            
QS$GW$                      & LDA           & FR & 3.25   & Ref.~\cite{Schilfgaarde_prb2021} \\
QS$G\hat{W}$                & LDA           & FR & 2.9    & Ref.~\cite{Schilfgaarde_prb2021} \\
\hline
\end{tabular}
\label{tab:gaps}
\end{table}
\noindent
Overall, the DMC estimate of $\Delta^{\rm DMC}_{f, \rm FR} = 2.9(1)$~eV shows good agreement with previous $GW$ results.
Although the $GW$ predictions are somewhat scattered  (2.59\cite{Wu_natcomm2019} to 3.25\cite{Schilfgaarde_prb2021} eV), our DMC value agrees much better with $GW$ results than with KS-DFT results, which are significantly lower.
In particular, our result agrees well with a previous QS$G\hat{W}$ result of 2.9 eV\cite{Schilfgaarde_prb2021}.

\begin{figure}[!htbp]
\centering
\includegraphics[width=0.50\textwidth]{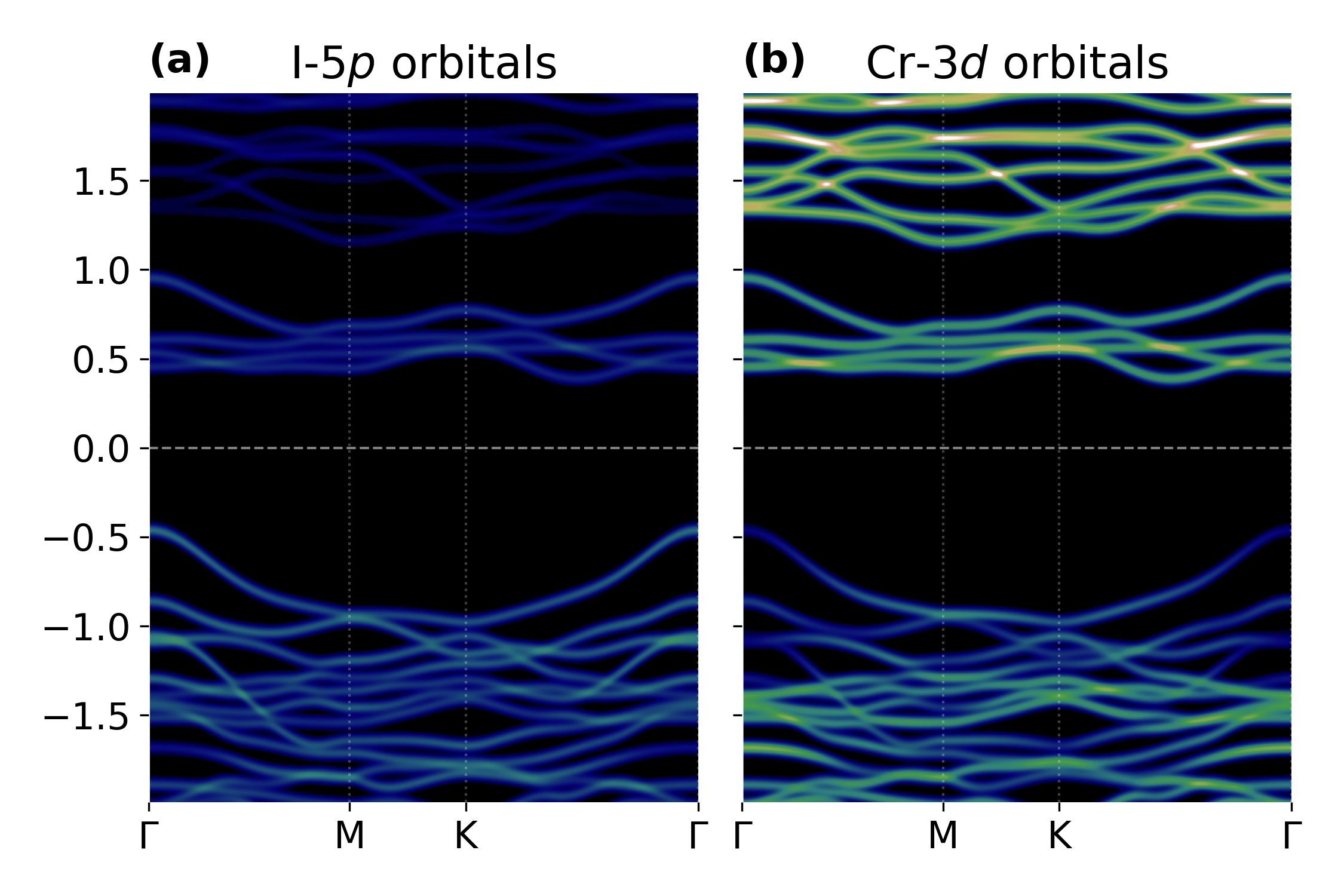}
\caption{
Fully-relativistic PBE band structures using projections to the atomic orbitals.
(a) I-$5p$ orbital and
(b) Cr-$3d$ orbital results are shown.
Brighter colors correspond to larger atomically-resolved densities.
}
\label{fig:fatbands}
\end{figure}

As of the writing of this paper, experimental data is available for ML CrI$_3$ on a sapphire substrate (we note that Ref. \onlinecite{Seyler_np2018} provides evidence that the optical properties of ML CrI$_3$ are minimally affected by substrate choice) using differential reflectance, which is proportional to absorbance\cite{Seyler_np2018}.
This optical data will show peaks for excitonic states as well as a peak for the fundamental gap. 
Our DMC gap $\Delta^{\rm DMC}_{f, \rm FR} = 2.9(1)$~eV best aligns with a sharp peak at $\approx 2.7$~eV.
This is the highest peak within the 3~eV window, but multiple wide peaks occur at lower energies than 2.7~eV corresponding respectively to two intra-atomic $d$-$d$ excitations (1.6, 1.9 eV), and two ligand-to-metal charge transfer (LMCT) excitations (2.2, 2.7 eV).\cite{Seyler_np2018} Although the LMCT excitation at $\approx 2.2$~eV was categorized as LMCT, it appears to have an excitonic character\cite{Wu_natcomm2019}, which we will not detect in our DMC calculations. In contrast, the $\approx2.7$~eV peak was experimentally categorized as a  LMCT excitation with the least excitonic character of the observed peaks, which would be consistent with how our trial wave functions were constructed.
Specifically, upon plotting the orbital-resolved band structure, Fig.~\ref{fig:fatbands}, we see that the VBM$\rightarrow$CBM transition should have an I-$5p$ to Cr-$3d$ character.
Namely, at the $\Gamma$-point, the VBM signal is larger for the I-$5p$ orbital, while the CBM signal is larger for the Cr-$3d$ orbital. Future experiments using angle-resolved photoemission spectroscopy (ARPES) on freestanding monolayer CrI$_3$ would be the best comparison for our DMC result.

The DMC value is slightly higher than the experimental value of 2.7~eV\cite{Seyler_np2018}.
This can be attributed to a few reasons, beyond sampling errors.
First, historically, the FNDMC fundamental gaps have been consistently predicted to be slightly larger than experimental gaps due to a larger fixed-node bias in the excited states compared to the ground states~\cite{meltonManybodyElectronicStructure2020, Annaberdiyev_prb2021}.
This is because the KS-DFT trial wave function is optimized for the ground state, and thus the virtual nodal surfaces are expected to be less accurate.
Second, both the $GW$ and KS-DFT methods predict that the SOC significantly reduces the gap.
For instance, $GW$ calculations on bulk CrI$_3$ predict a reduction by about 0.5~eV\cite{Kutepov_prm2021}.
Similarly, our PBE and PBE$+U_{2.0}$ KS-DFT calculations on ML CrI$_3$ predict a decrease of 0.38~eV and 0.5~eV, respectively (Table~\ref{tab:gaps}).
While we expect this gap reduction due to spin-orbit effects to be much less dramatic from our study, an extended study using FPSODMC with larger supercells and better sampling (currently out-of-scope of this paper) might reduce the $\sim$0.2 eV difference we currently see from experiments.

The single-reference FNDMC method has produced very successful results for the fundamental gaps of 2D materials\cite{wines_towards_2024}, such as phosphorene ($\Delta^{\rm {DMC}}_{f} = 2.53(2)$~eV\cite{frank_many-body_2019} vs. $\Delta^{\rm Exp}_{f} = 2.46$~eV\cite{gaufres_momentum-resolved_2019}), fluorographene ($\Delta^{\rm {DMC}}_{f} = 7.1(1)$~eV\cite{dubeckyFundamentalGapFluorographene2020} vs. $\Delta^{GW}_{f} = 7.14$~eV\cite{dubeckyFundamentalGapFluorographene2020}), and GeSe ($\Delta^{\rm {DMC}}_{f} = 1.62(16)$~eV\cite{Shin_PhysRevMats} vs. $\Delta^{\rm Exp}_{f} = 1.53$~eV\cite{wines_towards_2024}).
The result of this work is another case of obtaining accurate results for the excited state of a 2D material, which in this case is arguably more complicated due to the presence of transition metals and ferromagnetism.
Transition metals such as Cr often display multireference character (such as in the Cr$_2$ dimer\cite{larsson_chromium_2022}), for which the single-reference DMC method and other single-reference quantum chemistry methods such as CCSD(T) can be expected to struggle\cite{li_accurate_2020, purwanto_auxiliary-field_2015}.
Therefore, how can we justify the use of single-reference trial wave functions for the gap in this system?
It is not trivial to answer this question, as one has to show that the true excitation can be properly described by a VBM$\rightarrow$CBM single-electron promotion.

In an attempt to answer this question, we characterized the electronic transition based on the difference between the density matrices (DMs) of the ground ($\rho_\mathrm{GS}$) and excited states ($\rho_\mathrm{EX}$) in ML CrI$_3$, similar to the procedure described in Refs.~\cite{head-gordon_analysis_1995, shin_systematic_2024}.
To obtain accurate DMs, we have employed a configuration interaction (CI) method, CIPSI (Configuration Interaction using a Perturbative Selection done Iteratively)\cite{Garniron_jcp2017, garnironQuantumPackageOpenSource2019}.
We have run CIPSI and constructed wave functions with more than 5 million determinants at the $\Gamma$-point of the primitive cell.
The $\rho_\mathrm{GS}$ and $\rho_\mathrm{EX}$ DMs were constructed using these multideterminant wavefunctions to obtain the difference between these DMs:
\begin{equation}
    \label{eqn:dm_diff}
    \Delta \rho = \rho_\mathrm{EX} - \rho_\mathrm{GS} 
\end{equation}
\begin{equation}
    \label{eqn:diag}
    D = P^{-1} \Delta \rho P,
\end{equation}
where $P$ is a matrix with the eigenvectors of $\Delta \rho$ as columns.
The result is a diagonal matrix with positive and negative values, which correspond to electrons and holes of the excitation, respectively.
Namely, for a given orbital basis, the negative occupations represent the orbitals from which the charge is ``detached'' and the positive occupations represent the orbitals to which they ``attach,'' with the trace of $\Delta \rho$ giving a net zero.
The above procedure was done separately using PBE$+U_{2.0}$ orbitals and natural orbitals (NOs) in the multideterminant wave functions.
The NOs are useful for this analysis since they are considered optimal orbitals for CI because they provide the fastest convergence in expansions.
These NOs were obtained by diagonalizing the DMs corresponding to wave functions with more than 5 million determinants (see the SI for the methodological details\cite{Sun_wcms2018, Sun_jcp2020, Benali_jcp2020}).

\begin{figure}[!htbp]
\centering
\includegraphics[width=0.5\textwidth]{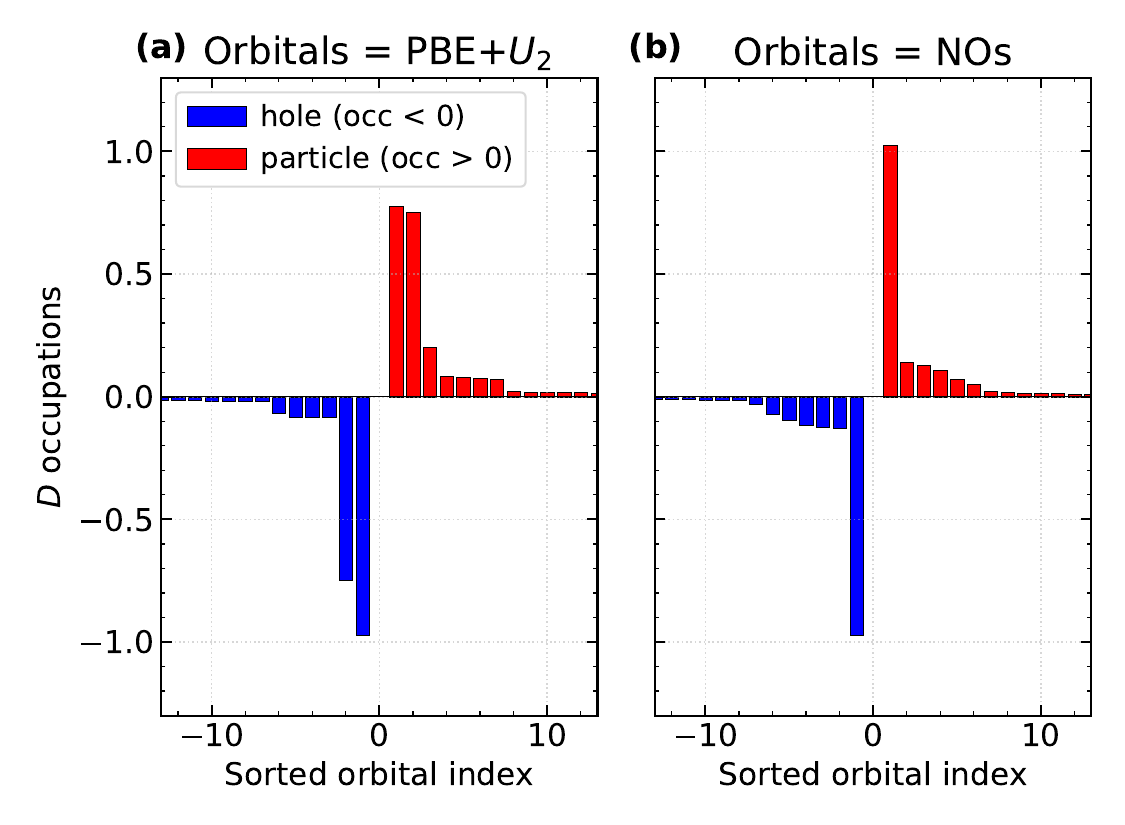}
\caption{
Diagonal occupations of the matrix $D$ (see Eqns.~\ref{eqn:dm_diff}, \ref{eqn:diag}).
(a) Initial single particle orbitals from PBE$+U_{2.0}$ eV SCF and (b) CIPSI-optimized single-particle natural orbitals (NOs).
Red denotes particle occupations (occ $>$ 0) and blue denotes hole occupations (occ $<$ 0).
}
\label{fig:cipsi_occ}
\end{figure}

\begin{figure}[!htbp]
\centering
\includegraphics[width=0.45\textwidth]{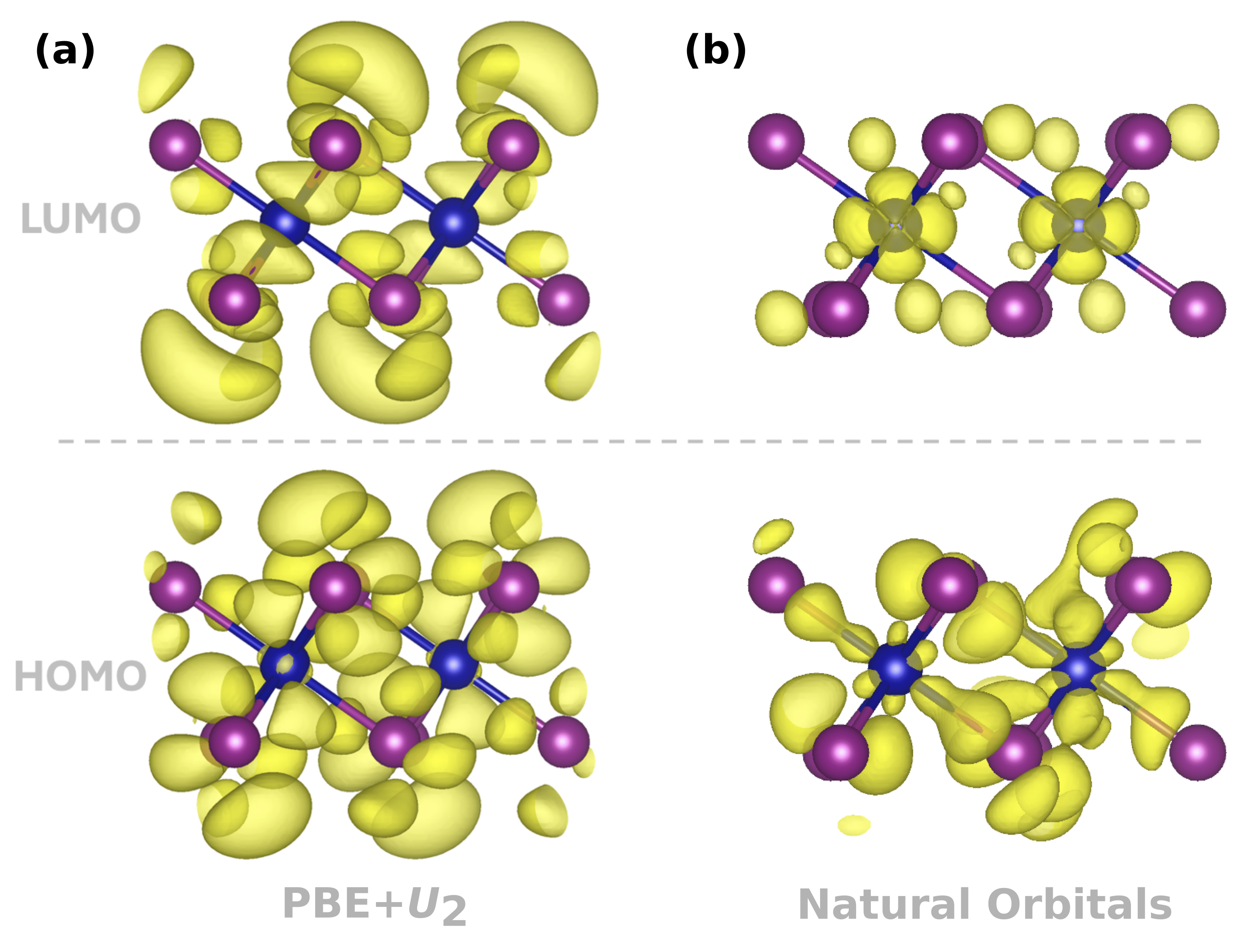}
\caption{
Monolayer CrI$_3$'s highest-occupied (HO) and lowest-unoccupied (LU) orbital density isosurfaces at the $\Gamma$-point.
(a) PBE+$U_{2.0}$ orbitals and
(b) natural orbitals are shown.
Note the more localized densities in the NOs case.
Center (blue) atoms are Cr and edge (purple) atoms are I.
}
\label{fig:iso}
\end{figure}

The results from the above procedure are shown in Fig.~\ref{fig:cipsi_occ}.
With the PBE$+U_2$ orbitals, the particle and the hole occupations are spread across a few orbitals, and the magnitude of those occupations is significantly smaller than those based on the NOs.
On the other hand, the particle and hole character with the NOs is dominated by a single orbital, with the trailing occupations much smaller than the dominant one.
In other words, the PBE$+U_2$ orbitals show smeared-out occupations while the NOs show sharp occupations defined by the highest occupied (HO) to lowest unoccupied (LU) orbital transition.
This suggests that the fundamental gap \textit{can} be described by a VBM$\rightarrow$CBM single-electron transition, given an appropriate orbital basis.
In this particular case, the NOs correspond to a more localized charge density, as shown in Fig.~\ref{fig:iso}.
In addition, note that the I-$5p$-to-Cr-$3d$ excitation character can be seen more easily in Fig.~\ref{fig:iso} with the NOs.

If the electronic transition corresponding to the fundamental gap \textit{did} require a multireference expansion, we would expect the particle or hole occupations in Fig.~\ref{fig:cipsi_occ}(b) to be spread across multiple orbitals, even with the use of NOs, which is the best possible basis for CI.
A considerable multireference character may still exist in the individual states, but we are interested in the ability of single-reference wavefunctions to describe the gap, rather than the total energies of each state. 

We note that, although our DMC calculations did not use NOs, the above test indicates that the excitation should be sufficiently described by a one-particle VBM$\rightarrow$CBM transition using a single reference wave function with an appropriate basis.
In FNDMC, the appropriate basis is dictated by the nodal hypersurface, and the best trial wave function can be selected variationally within a chosen parameter space that minimizes the energy.
In this case, we have used the Hubbard $U$ parameter, which was previously shown to result in the lowest energies for the ground states of the monolayer and bulk for a value close to $U = 2.0$~eV~\cite{Ichiba_prm2021,Staros_jcp2022,wines_systematic_2023}.
We have also tested that the PBE$+U_{2.0}$ trial wave function gives a significantly lower energy than the PBE one for the excited state ($E^\dag_{N}(\Gamma)$) for a $[2\times2\times1]$ supercell (see the SI).
Further comparisons and verification using the nodal surfaces of hybrid-DFT and CI NOs would be useful, although perhaps outside the scope of this work.

\section{Conclusions}
\label{sec:conclusions}

We have applied the high-accuracy fixed-node and fixed-phase spin-orbit Diffusion Monte Carlo (DMC) methods to predict the fundamental gap of monolayer (ML) CrI$_3$.
To the best of our knowledge, this is the first time these methods have been applied to the excited state of a magnetic 2D material. The DMC gap obtained of 2.9(1) eV agrees well with the highest peak in the optical reflectance spectrum (within a 3 eV window), which is located at $\approx 2.7$~eV. In addition, our DMC gap agrees well with previous $GW$ results, especially with a QS$G\hat{W}$ result of 2.9~eV~\cite{Schilfgaarde_prb2021}, while it disagrees with non-hybrid density functional theory (DFT).
This highlights the importance of using post-HF or post-DFT methods for 2D materials, such as $GW$, CI, or DMC.

In ML CrI$_3$, we showed that electron correlations are sufficiently captured with single-reference DMC to determine the fundamental gap.
This is corroborated by the agreement with the optical experimental results and by an orbital occupation analysis using natural orbitals of the selected configuration interaction method.
Moreover, our DMC calculations suggest that the band gap renormalization occurs mainly due to electron correlation effects, and the spin-orbit coupling plays a relatively minor role.
Specifically, the same DMC fundamental gap was obtained (within $\pm 0.1$~eV statistical errors) upon the inclusion of explicit spin-orbit effects, despite the underlying DFT method predicting a decrease due to spin-orbit coupling of half an eV.
Therefore, ML CrI$_3$ constitutes a case where accounting for electron correlations is critical, signifying the need for going beyond DFT in 2D materials with magnetism.

\section*{Associated Content}
The input and output files generated in this work were published in the Materials Data Facility~\cite{blaiszik_materials_2016, blaiszik_data_2019} and can be found in Ref.~\cite{mdf_data}.
See the SI for the employed structure, pseudopotential accuracy tests, and further details on our DFT, CIPSI, and QMC methodology.

\section*{Acknowledgements}

We thank P. R. C. Kent for reading the manuscript and providing helpful suggestions.

This work was supported by the U.S. Department of Energy, Office of Science, Basic Energy Sciences, Materials Sciences and Engineering Division, as part of the Computational Materials Sciences Program and Center for Predictive Simulation of Functional Materials.
D.S. was initially supported by the U.S. Department of Energy through the Office of Science Graduate Student Research (SCGSR) Program.  Part of this research was conducted as part of a user project at the Center for Nanophase Materials Sciences (CNMS), which is a US Department of Energy, Office of Science User Facility at Oak Ridge National Laboratory.  This research used resources of the Compute and Data Environment for Science (CADES) at the Oak Ridge National Laboratory, which is supported by the Office of Science of the U.S. Department of Energy under Contract No. DE-AC05-00OR22725"

An award of computer time was provided by the Innovative and Novel Computational Impact on Theory and Experiment (INCITE) program.
This research used resources of the Oak Ridge Leadership Computing Facility, which is a DOE Office of Science User Facility supported under Contract No. DE-AC05-00OR22725.
This research used resources of the Argonne Leadership Computing Facility, which is a DOE Office of Science User Facility supported under contract DE-AC02-06CH11357.
This research used resources of the National Energy Research Scientific Computing Center (NERSC), a U.S. Department of Energy Office of Science User Facility located at Lawrence Berkeley National Laboratory, operated under Contract No. DE-AC02-05CH11231. We gratefully acknowledge the computing resources provided on IMPROV, a high-performance computing cluster operated by the Laboratory Computing Resource Center (LCRC) at Argonne National Laboratory.

This paper describes objective technical results and analysis. Any subjective views or opinions that might be expressed in the paper do not necessarily represent the views of the U.S. Department of Energy or the United States Government.

\section*{Competing Interests}
The authors declare no competing financial or non-financial interests.

\section*{Author Contributions}
D.S. (DFT, DMC, CIPSI calculations and writing).
K.G. and A.B. (CIPSI calculations and QP2 code developments).
A.A. (DFT, DMC calculations and writing).
P.G. and B. R (Project initiation and supervision).
All authors read, edited, and approved the final manuscript.

\textsuperscript{\textdagger} indicates equal contribution as a first author. 
\section*{References}
\bibliography{main}

\makeatletter\@input{xx.tex}\makeatother
\end{document}

\include{MyCommand}
\setlength{\tabcolsep}{5pt} % for the horizontal padding
\renewcommand{\arraystretch}{1.15}
\renewcommand{\thetable}{\arabic{table}}

\title{Supplementary Material for: ``A many-body characterization of the fundamental gap in monolayer CrI$_3$''}

\author{Daniel Staros}
\affiliation{Department of Chemistry, Brown University, Providence, Rhode Island 02912, USA}
\author{Abdulgani Annaberdiyev}
\affiliation{\mbox{Center for Nanophase Materials Sciences Division, Oak Ridge National Laboratory, Oak Ridge, Tennessee 37831, USA}}
\author{Kevin Gasperich}
\affiliation{\mbox{Computational Science Division, Argonne National Laboratory, Lemont, Illinois 60439, USA}}
\author{Anouar Benali}
\affiliation{\mbox{Computational Science Division, Argonne National Laboratory, Lemont, Illinois 60439, USA}}
\author{Panchapakesan Ganesh}
\affiliation{\mbox{Center for Nanophase Materials Sciences Division, Oak Ridge National Laboratory, Oak Ridge, Tennessee 37831, USA}}
\author{Brenda Rubenstein}
\email{Authors to whom correspondence should be addressed: Brenda Rubenstein, brenda\_rubenstein@brown.edu and Ganesh Panchapakesan, gpanchapakesan@ornl.gov}
\affiliation{Department of Chemistry, Brown University, Providence, Rhode Island 02912, USA}

%\date{\today}
\begin{abstract}
The Supplementary Material provides the employed monolayer CrI$_3$ geometry (Section~\ref{sec:geom}), pseudopotential accuracy tests (Section~\ref{sec:ecp}), DMC total energies and gaps (Section~\ref{sec:dmc}), and CIPSI methodology details (Section~\ref{sec:cipsi}).
\end{abstract}

%\pacs{}
\maketitle
%\pagenumbering{arabic}

\section{Monolayer CrI$_3$ Geometry}
\label{sec:geom}

\begin{table}[!htbp]
\centering
\caption{
The employed primitive cell geometry of monolayer CrI$_3$ in \texttt{xsf} format (Angstrom units).
}
\label{tab:geom_xsf}
\begin{tabular}{lrrrrr}
\hline
\multicolumn{2}{l}{CRYSTAL} \\
\multicolumn{2}{l}{PRIMVEC} \\
   &  6.84482176 &  0.00000000 &  0.00000000 \\
   & -3.42241088 &  5.92778953 &  0.00000000 \\
   &  0.00000000 &  0.00000000 & 26.74842620 \\
\multicolumn{2}{l}{PRIMCOORD} \\
8 1 \\
53 &  4.44148330 &  0.00000000 & 14.93935014 \\
53 &  1.20166923 &  2.08135216 & 14.93935014 \\
53 & -2.22074165 &  3.84643737 & 14.93935014 \\
53 &  2.40333845 &  0.00000000 & 11.80907605 \\
53 &  2.22074165 &  3.84643737 & 11.80907605 \\
53 & -1.20166923 &  2.08135216 & 11.80907605 \\
24 & -0.00000000 &  3.95185968 & 13.37421310 \\
24 &  3.42241088 &  1.97592984 & 13.37421310 \\
\hline
\end{tabular}
\end{table}

\section{Pseudopotential Accuracy}
\label{sec:ecp}

\begin{table}[htbp!]
\centering
\caption{Estimate of the pseudopotential accuracy for the fundamental gap ($\Delta_f$) calculations.
``Soft'' represents the pseudopotential set used in the main article: RRKJ for Cr~\cite{krogel_pseudopotentials_2016} and BFD for I~\cite{burkatzki_energy-consistent_2007}.
``ccECP'' are harder pseudopotentials optimized using correlated calculations~\cite{kincaidCorrelationConsistentEffective2022, wang_new_2022}.
Scalar-relativistic pseudopotentials were used for these calculations.
All values are in eV.
}
\begin{tabular}{lccr}
\hline
Method      & Soft $\Delta_f$ & ccECP $\Delta_f$ & $\delta$ $\Delta_f$ \\
\hline
PBE         & 1.289           & 1.323            & -0.033 \\
PBE0        & 1.769           & 1.796            & -0.027 \\
\hline
\end{tabular}
\label{tab:pp_errors}
\end{table}

\clearpage
\section{DMC Energies}
\label{sec:dmc}

\begin{table}[htbp!]
\centering
\caption{
Scalar-relativistic FNDMC total energies [Ha] per formula unit $N_{\rm f.u.}$.
$E_{N}(\Gamma)$ is the ground state, $E^\dag_{N}(\Gamma)$ is the excited state,  $E_{N-1}(\Gamma)$ is the cation, and $E_{N+1}(\Gamma)$ is the anion energy.
$\infty$ represents the thermodynamic limit using a $N_{\rm f.u.} = [8, 18]$ two-point extrapolation.
PBE$+U_{2.0}$ trial wave function was used for all values in this table. 
}
\begin{tabular}{cllll}
\hline
$N_{\rm f.u.}$ & $E_{N}(\Gamma)$ & $E^\dag_{N}(\Gamma)$ & $E_{N-1}(\Gamma)$ & $E_{N+1}(\Gamma)$ \\
\hline
2        & -121.1351(1) & -121.1043(1) & -121.0198(1) & -121.1295(1) \\
4        & -121.1203(2) & -121.1011(2) & -121.0719(2) & -121.1270(3) \\
6        & -121.1155(4) & -121.1021(2) & -121.0853(3) & -121.1219(8) \\
8        & -121.1158(4) & -121.1055(2) & -121.0938(3) & -121.1213(3) \\
18       & -121.1167(2) & -121.1115(1) & -121.1076(2) & -121.1192(2) \\
\hline
$\infty$ & -121.11744   & -121.11626   & -121.11873   & -121.11742   \\
\hline
\end{tabular}
\label{tab:qmc_totals}
\end{table}

\begin{table}[htbp!]
\centering
\caption{
Fully-relativistic FPSODMC total energies [Ha] per formula unit $N_{\rm f.u.}$.
$E_{N}(\Gamma)$ is the ground state, $E^\dag_{N}(\Gamma)$ is the excited state,  $E_{N-1}(\Gamma)$ is the cation, and $E_{N+1}(\Gamma)$ is the anion energy.
PBE$+U_{2.0}$ trial wave function was used for all values in this table. 
}
\begin{tabular}{cllll}
\hline
$N_{\rm f.u.}$ & $E_{N}(\Gamma)$ & $E^\dag_{N}(\Gamma)$ & $E_{N-1}(\Gamma)$ & $E_{N+1}(\Gamma)$ \\
\hline
8 & -121.0534(2) & -121.0429(3) & -121.0334(4) & -121.0574(2) \\
\hline
\end{tabular}
\label{tab:qmc_tot_fr}
\end{table}

\begin{table}[htbp!]
\centering
\caption{
DMC gaps [eV] for different supercell sizes with $N_{\rm f.u.}$ formula units.
$\infty$ represents the thermodynamic limit using a $N_{\rm f.u.} = [8, 18]$ two-point extrapolation.
$\Delta^{\rm avg}_{f}(\Gamma) = [\Delta^{\rm qp}_{f}(\Gamma) + \Delta^{\rm opt}_{f}(\Gamma)]/2$ is the average of two definitions.
Scalar-relativistic (FNDMC, SR) and fully-relativistic (FPSODMC, FR) values are shown.
PBE$+U_{2.0}$ trial wave function was used for all values in this table. 
}
\begin{tabular}{c|lll|lll}
\hline
$N_{\rm f.u.}$ & $\Delta_{f, \rm SR}^{\rm qp}(\Gamma)$ & $\Delta_{f, \rm SR}^{\rm opt}(\Gamma)$ & $\Delta_{f, \rm SR}^{\rm avg}(\Gamma)$ & $\Delta_{f, \rm FR}^{\rm qp}(\Gamma)$ & $\Delta_{f, \rm FR}^{\rm opt}(\Gamma)$ & $\Delta_{f, \rm FR}^{\rm avg}(\Gamma)$ \\
\hline
2        & 6.581(19) & 1.678(10) & 4.129(13) & \\
4        & 4.549(54) & 2.098(30) & 3.324(37) & \\
6        & 3.88(20)  & 2.185(80) & 3.03(13)  & \\
8        & 3.59(21)  & 2.24(10)  & 2.91(15)  & 3.49(14) & 2.283(80) & 2.886(94) \\
18       & 3.25(21)  & 2.56(10)  & 2.90(14)  & \\
\hline
$\infty$ & 2.970     & 2.815     & 2.892     \\
\hline
\end{tabular}
\label{tab:qmc_gaps}
\end{table}

\begin{table}[htbp!]
\centering
\caption{
Scalar-relativistic FNDMC total energies [Ha] per formula unit $N_{\rm f.u.}$ for PBE and PBE$+U_{2.0}$ trial wave functions.
$E_{N}(\Gamma)$ is the ground state, $E^\dag_{N}(\Gamma)$ is the excited state,  $E_{N-1}(\Gamma)$ is the cation, and $E_{N+1}(\Gamma)$ is the anion energy.
}
\begin{tabular}{llllll}
\hline
XC/Orbital & $N_{\rm f.u.}$ & $E_{N}(\Gamma)$ & $E^\dag_{N}(\Gamma)$ & $E_{N-1}(\Gamma)$ & $E_{N+1}(\Gamma)$ \\
\hline
PBE             & 8 & -121.1158(3) & -121.1041(2) & -121.0936(2) & -121.1209(3) \\
PBE$+U_{2.0}$   & 8 & -121.1158(4) & -121.1055(2) & -121.0938(3) & -121.1213(3) \\
\hline
\end{tabular}
\label{tab:ref_comp}
\end{table}

\clearpage
\section{CIPSI Methods}
\label{sec:cipsi}

As the first step in our CIPSI orbital optimization, we self-consistently converge the primitive cell of ML CrI$_3$ using the open-source \texttt{PySCF} program~\cite{Sun_wcms2018, Sun_jcp2020}.
This calculation is done at the $\Gamma$ point using a vTZ basis of periodic Gaussian-based atomic orbitals.
The primitive exponents were tuned such that the basis set error is about 7 mHa relative to an identical, fully-converged plane-wave relaxation.
We use the PBE+$U$ exchange-correlation functional with a Hubbard $U$ correction of 2 eV per Cr in accordance with the \texttt{Quantum Espresso} calculations.
These orbitals form the starting point for our iterative CIPSI orbital optimization.

\begin{figure}[!htbp]
\centering
\begin{subfigure}{0.5\textwidth}
\includegraphics[width=\textwidth]{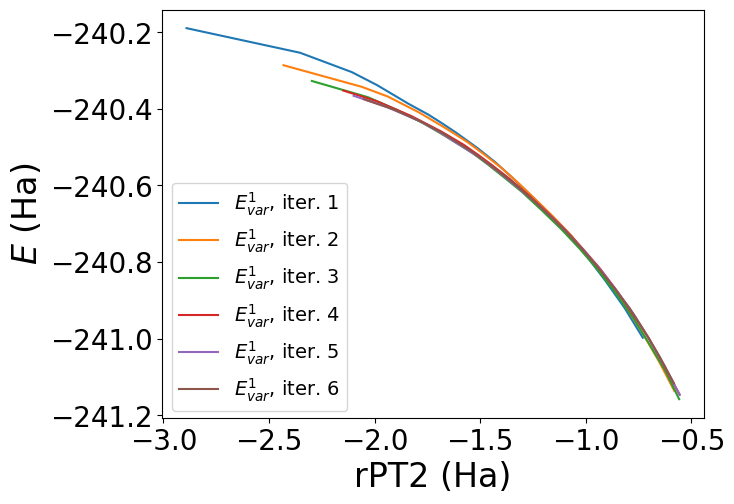}
\caption{Ground State}
%\label{fig:}
\end{subfigure}%
\begin{subfigure}{0.5\textwidth}
\includegraphics[width=\textwidth]{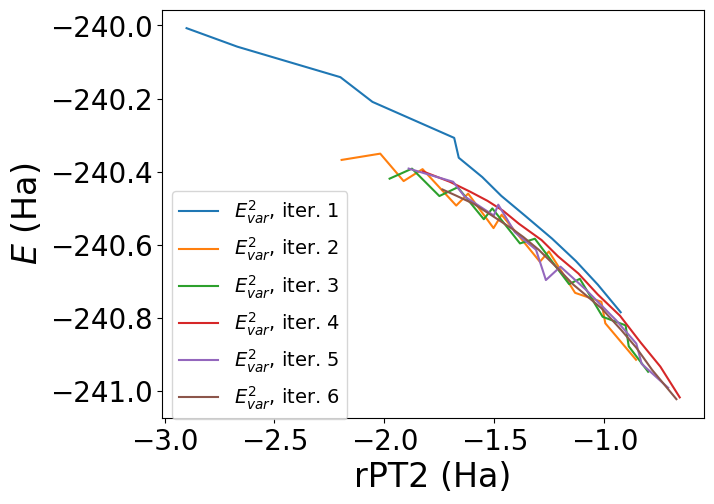}
\caption{Excited State}
%\label{fig:}
\end{subfigure}
\caption{
The convergence of natural orbitals in ML CrI$_3$ using repeated CIPSI runs.
The variational energies of CIPSI vs. rPT2 are shown for (a) the ground and (b) the first excited state.
Smaller perturbative corrections (rPT2) correspond to expansions with larger numbers of determinants.
Both sets of iterations were performed starting from the same set of restricted PBE+$U_{2.0}$ orbitals.
Note that the $N_{\rm iter} \geq 3$ curves closely overlap with each other, signifying the convergence of natural orbitals.
}
\label{fig:cipsi_conv}
\end{figure}

To obtain the natural orbitals of monolayer CrI$_3$, we start by performing a CIPSI expansion~\cite{Huron_jcp1973} around the self-consistently converged DFT determinant.
In this process, the most important determinants $\alpha$ in this expansion are iteratively selected for both the ground and first excited states to favor those with larger estimated contributions to the electronic correlation energy as determined by a weighted sum $e_{\alpha}^{(n)}$ of their corresponding second-order perturbative corrections to each state $\ket{\Psi_{i}^{(n)}}$ for the $n^{\textnormal{th}}$ iteration:\cite{Garniron_jcp2017} 
\begin{equation}
    e_{\alpha}^{(n)} = \sum_{i=1}^{N_\textrm{states}} w_i\frac{|\braket{\Psi_{i}^{(n)}|\hat{H}|\alpha}|^2}{\braket{\Psi_{i}^{(n)}|\hat{H}|\Psi_{i}^{(n)}}-\braket{\alpha|\hat{H}|\alpha}}
\end{equation}
\noindent
We used equal weights $w_i$ for each state $i$, and a convergence criterion for the total perturbative corrections $E_{PT2}^{(n)}$ is enforced.
Next, one-body density matrices of the corresponding expansions are constructed and diagonalized to yield a single determinant of NOs~\cite{Benali_jcp2020}.
Another such expansion is then done around this single determinant of NOs, which are transformed and diagonalized.
This forming of NOs and running CIPSI is iterated (6 times) until convergence is achieved.
We perform this procedure in two different ways: (i) separately for the ground state and first excited state ($E^\dag_{N}(\Gamma)$) to facilitate our calculation of the optical gap, and (ii) together for both states via the simultaneous selection of determinants that contribute to both states.
In the case of (i), this procedure is repeated until the variational energy as a function of renormalized perturbative corrections (rPT2) no longer changes with additional iterations.
This is shown in Fig.~\ref{fig:cipsi_conv}.
We find that it is currently challenging to fully converge the energies in the case of (ii).

Throughout this procedure, our active space consists of a total of 172 occupied and virtual orbitals, and the unrestricted DFT starting point consists of 38 up- and 32 down-spin electrons since there are nominally 3 unpaired up-spin electrons per Cr atom.
Each expansion is done within the Hilbert subspace of the first ten million determinants and targets the ground and first excited states, which preserve the spin multiplicity of $2S+1 = 7$.

All \texttt{PySCF} and consequent CIPSI calculations employ the correlation-consistent effective core potentials (ccECPs)~\cite{kincaidCorrelationConsistentEffective2022, wang_new_2022}.

\clearpage
\section*{References}
\bibliography{main}